# Countermeasure against blinding attacks on low-noise detectors with background noise cancellation scheme

Min Soo Lee, Byung Kwon Park, Min Ki Woo, Chang Hoon Park, Yong-Su Kim, Sang-Wook Han*, and Sung Moon

*Abstract*— **We developed a countermeasure against blinding attacks on low-noise detectors with a background noise cancellation scheme in quantum key distribution (QKD) systems. Background noise cancellation includes self-differencing and balanced avalanche photon diode (APD) schemes and is considered a promising solution for low-noise APDs, which are critical components in high-performance QKD systems. However, its vulnerability to blinding attacks has been recently reported. In this work, we propose a new countermeasure that prevents this potential security loophole from being used in detector blinding attacks. An experimental QKD setup is implemented and various tests are conducted to verify the feasibility and performance of the proposed method. The obtained measurement results show that the proposed scheme successfully detects occurring blinding-attack-based hacking attempts.**

## I. INTRODUCTION

Quantum key distribution (QKD) systems are being actively studied and pursued [1–13]. In the commercial arena, in particular, great progresses have been recently achieved in the area of high-speed QKD and long distance quantum network systems [14–25]. Many studies on QKD security have been also conducted uninterruptedly since 1984, because providing security is essential for the commercialization of such systems. Many research groups have focused on quantum hacking and countermeasures against it, to overcome loopholes resulting from the imperfections of hardware devices. In fact, attempts to

This work was supported by the ICT R&D programs of MSIP/IITP (Grants B0101-16-0060 and B0101-16-1355), the Ministry of Science, ICT and Future Planning Program (Grant 2N41850), and the KIST Research Program (Grant 2E26681).

Min Soo Lee and Byung Kwon Park are with the Center for Quantum Information, Korea Institute of Science and Technology, Seoul 136-791, Korea, and with the Department of Nano-materials, Science and Engineering, Korea University of Science and Technology, Daejeon 305-350, Korea (e-mail: lms0408@kist.re.kr; bkpark@kist.re.kr).

Chang Hoon Park is with the Center for Quantum Information, Korea Institute of Standards and Technology, Seoul 136-791, Korea, and with the Department of Electrical and Computer Engineering, University of Ajou, Suwon 443-749, Korea (e-mail: originalpch@kist.re.kr).

Min Ki Woo, Yong-Su Kim, Sang-Wook Han, and Sung Moon are with the Center for Quantum Information, Korea Institute of Standards and Technology, Seoul 136-791, Korea (e-mail: 022685@kist.re.kr; yong-su.kim@kist.re.kr; swhan@kist.re.kr; s.moon@kist.re.kr).

*Corresponding author.

hack avalanche photon diodes (APDs) have posed real threats to commercial QKD systems, although there have been many applicable hacking methods including blinding, time shifting, detector dead time, phase remapping, Trojan horses, Faraday mirrors, wavelength, phase information, and device calibration attacks [26–41]. In the research area of avalanche photon diodes, there has been remarkable progress in terms of higher speed and lower noise operations. In particular, avalanche photon diodes with background noise cancellation (BNC) schemes which include self-differencing and balanced APDs have been highlighted, because of their superior characteristics of GHz-speed operation and lower after-pulse noise [42–47]. It has been reported that QKD systems with self-differencing avalanche photon diodes can also be hacked [48]. However, threats to QKD systems with balanced APDs have not been mentioned yet. In this paper, we show a hacking method applicable to balanced APDs and propose a countermeasure scheme against such quantum hacking. In the following sections, we will discuss the concept of quantum hacking and its countermeasures. An experimental setup for verification will then be presented. Finally, some test results and the corresponding analysis will be discussed at length. In addition, we will briefly note that our countermeasure can also be used in self-differencing schemes.

## II. BLINDING SINGLE PHOTON DETECTORS TO ENABLE QUANTUM HACKING

As mentioned above, APDs with a BNC scheme constitute a promising solution for QKD systems, because of their superior noise characteristics. Normally, APDs are affected by two types of noise: dark count and after-pulse noises. Cooling the APD can easily reduce the dark count noise; however, this results in an increase of the after-pulse noise. Therefore, specific techniques are required to reduce the after-pulse noise while maintaining the same (or less) dark count noise. From this point of view, the BNC scheme is currently considered the most efficient technique for reducing the after pulse noise. However, it has been disclosed that the scheme is susceptible to blinding attacks. Having been originally designed to remove background noise, the BNC scheme removes both successive and simultaneous avalanche signals alike. The cancellation of avalanche signals means that an APD can be easily blinded.

Fig. 1 demonstrates one of the BNC schemes (balanced APD), and shows how it can be attacked using its avalanche signal cancelation behavior. In this scheme, the output

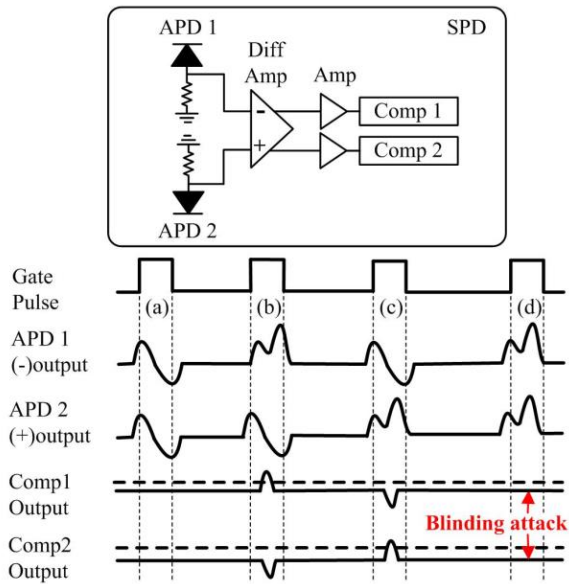

FIG. 1. Block diagram and timing chart of the balanced APD method with an attempted blinding attack.

APD: Avalanche photo diode; Diff Amp: Differential amplifier; Amp: Amplifier; Comp: Comparator; SPD: Single photon detector.

difference between two avalanche photo diodes (APDs) is delivered to the comparators (Comp), so that the background noise is cancelled, as shown by pulse (a) in Fig. 1. Normally, amplifiers are located between Diff amp and Comp to adjust the amount of background noise signals from two APDs so that two signals have similar value. This adjustment scheme do not effect on the performance of APDs and the attack threat by Eve. If an avalanche signal is generated only in APD 1, positive and negative signals occur in Comp 1 and Comp 2, respectively [Fig. 1, pulse (b)]. In the similar way, if an avalanche signal is generated only in APD2, negative and positive signal occur in Comp 1 and Comp 2, respectively [Fig.1, pulse (c)]. In both cases, the APD can correctly detect an avalanche signal with low background noise. However, if an avalanche signal is generated simultaneously in both APDs, the two avalanche signal contributions will cancel each other, resulting in APD blinding [Fig. 1, pulse (d)]. Therefore, an eavesdropper (Eve) can easily blind the APD by applying strong laser pulses to both APDs simultaneously. By blinding the APDs, Eve can herself control the QKD system without increasing the quantum bit error rate (QBER), and thus succeed in hacking the system. In the next section, we discuss how to hack practical plug and play (P&P) QKD systems using an intercept-and-resend method [27].

## III. Quantum Hacking of Plug and Play QKD Systems With a Balanced APD Scheme

 The intercept-and-resend is one of the critical QKD hacking methods. It is well known that, with this method, Eve can only get full secret keys from practical QKD systems if she blinds and controls the detectors [27]. The concept of an intercept-and-resend method on a P&P QKD system is shown in Fig. 2. The concept involves four different entities: Alice,

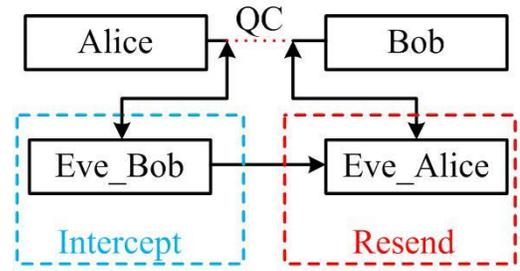

FIG. 2. Block diagram of an intercept-and-resend method on a plug and play QKD system.

QC: Quantum channel.

Bob, Eve–Bob, and Eve–Alice. Eve–Bob is designed to intercept the phase information from Alice and allow Eve to resend a single photon with modified phase from Eve–Alice to Bob. The attack proceeds as follows:

1) Bob generates and sends strong laser pulses to Alice. Before the arrival of these strong laser pulses to the Eve–Alice phase modulator, Eve–Bob generates strong laser pulses to Alice.

2) Alice randomly modulates (0, π/2, π, 3π/2) the phase of the strong laser pulses generated by Eve–Bob, and the modulated laser pulses return back to Eve–Bob (the *intercept* part). The phases (0, π) belong to basis 0 and (π/2, 3π/2) do to basis 1.

3) Eve–Bob modulates 0 (basis 0) or π /2 (basis 1) randomly and guesses the phase information of Alice using the click information from Eve–Bob's APDs.

4) If Alice and Eve-Bob have same bases, Eve can have correct phase information of Alice. On the other hand, if they have different bases, Eve can get only 50% correct phase information because an avalanche occurs in one of the APDs (randomly determined, with a 50 % probability). However, Eve guesses the phase information based on only Eve-Bob's APD clicks. And the 50% error is eliminated through sifting process and Eve's blinding attack as described in step 7), 8) and 9).

5) Eve-Bob transmits the guessed phase information to Eve-Alice, and Eve-Alice then modulates this phase information when Bob's original signal arrives at Eve-Alice. The modulated laser pulses will be resent to Bob, to generate APD clicks on Bob (the *resend* part).

6) Eve-Alice's modulated basis will either equal or differ from Bob's modulated basis. If equality is achieved, Eve can obtain all the secure keys from the P&P QKD system without increasing the QBER. If, however, the modulated bases differ, the APD clicks on Bob become error counts, and the QBER is therefore increased.

7) When Bob's APDs click, Alice, Bob, and Eve can obtain raw keys based on basis or APD click information. Alice

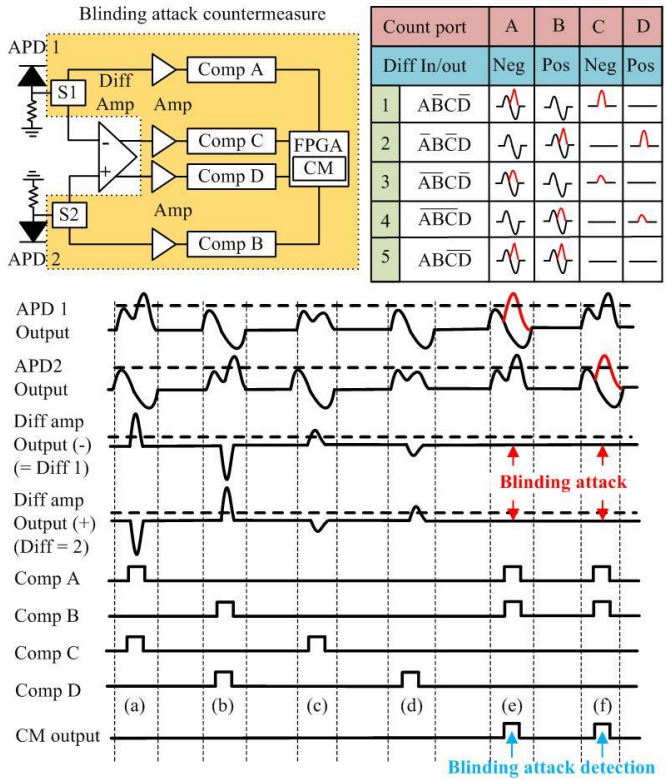

FIG. 3. Block diagram and timing chart of the countermeasure scheme against blinding attacks in QKD systems with a balanced APD scheme.

and Bob will then try to do a sifting processing through the classical channel. If the two parties have different bases, each part discards the raw keys. Eve hacks into the classical channel and also discards the raw keys (as Alice and Bob do) during the sifting processing. As a result, they obtain equally sifted keys, which will be the base of the secure keys in P&P QKD systems.

8) After the sifting process, Eve still has a major problem to solve, in the intercept-and-resend method. If the guessed phase information, based on Eve-Alice's basis, agrees with Bob's basis, all the APD clicks become sifting keys. However, Eve cannot know for sure which APD clicks become sifting keys and which become error counts. For example, if Eve-Bob's guessed phase is equal to Bob's phase, there will be a correct sifting key. However, if the guessed phase is wrong, there is an error count with a 50 % probability. As a result, Eve can be easily exposed by the increase of QBER caused by such error counts.

9) However, if Eve uses the discussed blinding attack method to the P&P QKD system, Bob's APDs will be blind and cannot generate any APD clicks; this means that there will be no QBER increase when Eve-Bob's guessed phase is different from Bob's phase. By providing not single photon but many photons from Eve to Bob, Eve can blind only when the guessed phase is different. If the guessed phase is the same, the APDs operate properly. As a result, Eve can control Bob's APDs to generate APD clicks only when Eve-Bob's and Bob's phase are equal.

This means that Eve can successfully obtain all the sifting keys without an increase in QBER. Plug and play systems using low noise APDs therefore need a new countermeasure scheme to thwart APD blinding attacks.

## IV. COUNTERMEASURE SCHEME

Fig. 3 shows the proposed countermeasure (CM) scheme, designed to thwart blinding attacks on QKD systems with a balanced APDs. The major purpose of the CM scheme is to detect the double click of APD 1 and 2 that blinds the APDs. Weak avalanches and strong avalanches are also detected and distinguished, by using logic gates on a field-programmable gate array (FPGA). A strong avalanche means that the output signal is larger than the background noise. If the output avalanche signal is smaller than the background noise, we call it a weak avalanche. The countermeasure consists of two splitters (S1 and S2), four comparators (Comp A, Comp B, Comp C, and Comp D), and an FPGA implementing the countermeasure scheme logic. Each APD (APD 1 and APD 2) is connected to one splitter (S1 and S2). Each splitter has one port to connect it to either Comp A or Comp B, and a second port to connect it to either the positive or negative input of the differential amplifier (see Fig. 3). The negative and positive outputs of the differential amplifier (Diff amp) are connected to Comp C and Comp D, respectively. Comp A and Comp B can detect strong avalanches in APD 1 or APD 2, and the differential amplifier evaluates the difference between the APD 1 and APD 2 output signals. If an avalanche occurs in APD 1, an output signal is generated in Comp C; conversely, if an avalanche signal occurs in APD 2, a signal is generated in Comp D. The CM is designed to detect attempted blinding attacks using simultaneous weak and strong avalanche signals. As shown in Fig. 3 [pulse (a)], if a strong avalanche signal occurs in APD 1, click signals occur in Comp A and Comp C, a situation that can be detected by a gate implementing the logic expression $A\overline{B}C\overline{D}$. In contrast, if click signals are generated in both Comp B and Comp D, this means that a strong avalanche signal has occurred in APD 2 [Fig. 3, pulse (b)], which can be detected by a logic gate implementing $\overline{A}B\overline{C}D$. When a weak avalanche signal is generated in APD 1, only one click signal occurs, in Comp C [Fig. 3, pulse (c)]; it can be detected by a logic gate implementing $\overline{A}B C\overline{D}$. When a click signal occurs only in Comp D, it means that a weak avalanche signal was generated in APD 2 [Fig. 3, pulse (d)]. This case can be detected by a logic gate implementing $\overline{A}B\overline{C}D$. If avalanche signals are generated in APD 1 and APD 2 simultaneously, there is no signal in either Comp C or Comp D. This means that the avalanche signals were obscured by the blinding signals generated by Eve [Fig. 3, pulses (e) and (f)]. This case can be detected by a logic gate implementing $AB\overline{C}\overline{D}$, and indicates that an attempted blinding attack has occurred. Therefore, if the output of this logic gate is evaluated, a CM output signal can be generated by the FPGA to disable raw key generation, and thus prevent the attempted quantum hacking.

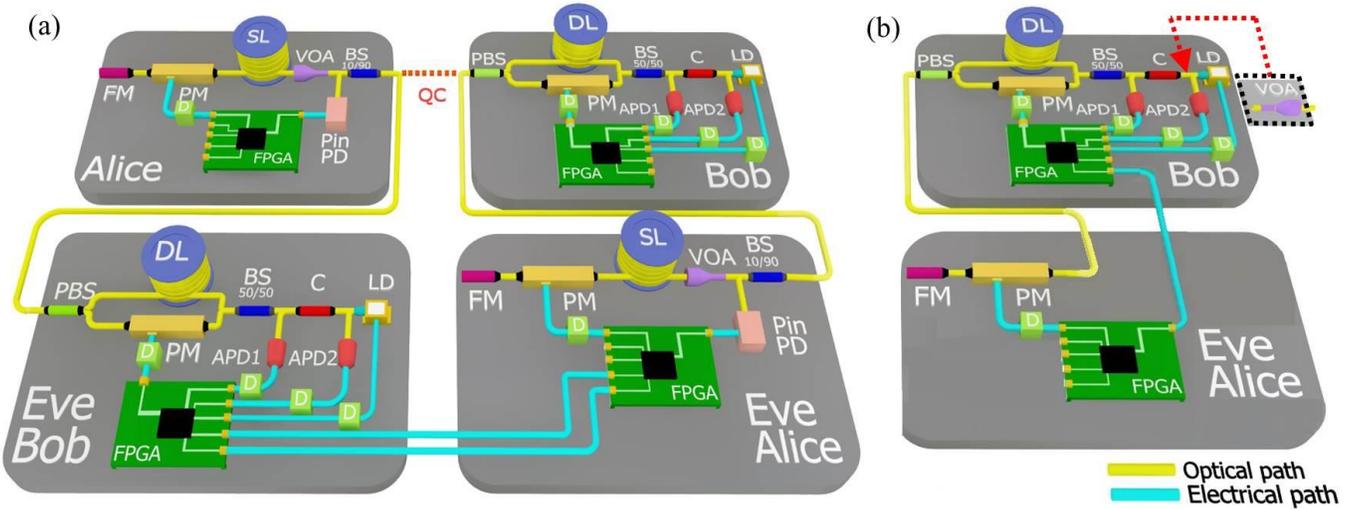

FIG. 4.   (a) Conceptual diagram of the experimental setup for an intercept-and-resend method on a plug and play QKD system. (b) Proof-of-principle experimental setup for a blinding attack.

FM: Faraday mirror; PM: Phase modulator; FPGA: Field-programmable gate array; SL: Storage line; VOA: Variable passive attenuator; PinPD: Pin photodiode; BS 90/10: Beam splitter 90:10; PBS: Polarization beam splitter; DL: Delay line; BS 50/50: Beam splitter 50:50; C: Circulator; LD: Laser diode; APD: Avalanche photodiode; PMC: Phase modulation controller; D: Driver (such as a PPG or NPG).

## V.  EXPERIMENTS AND RESULTS

The conceptual diagram of the setup to simulate an intercept-and-resend method on a practical P&P QKD is shown in Fig. 4(a). As discussed above, four entities are involved: Alice, Bob, Eve–Bob, and Eve–Alice. Eve–Bob is designed to incept a single photon that will be transmitted from Alice to Bob. In the intercept part of the simulated attack, Eve-Bob communicates with Alice as the same way of a conventional plug and play QKD system. In the resend part, Eve–Alice is designed to resend strong laser pulses for blinding attack after phase modulation as Eve–Bob's guessing phase modulation information. As a result, Eve–Alice can control Bob's APD click.

We prove the performance of the new proposed countermeasure against blinding attacks on low-noise APDs through a proof-of-principle experimental setup composed of only Eve–Alice and Bob, as shown in Fig. 4(b). We incorporate all cases of phase modulation in the simplified Eve–Alice, instead of using the full intercept part setup, which includes also Eve–Bob and Alice, as shown in Fig. 4(a). The phase modulation timing is controlled by Bob's FPGA; a driver (D) modulates the laser pulse phase. A VOA is located on Bob for experimental convenience. Bob's FPGA (Altera Stratix 4) produces a 2 MHz clock signal to trigger a negative pulse generator (NPG; Avteck AVP-AV-1) used to drive a laser diode, and a positive pulse generator (PPG; Agilent 81160A) to drive the APDs (Princeton Lightwave PGA-308). The PPG generates periodic rectangular voltage pulses with 5-V amplitude and 2-ns width. Using a bias tee (Avteck AVX-T), the rectangular pulses are superimposed on a DC offset voltage, to operate the APD in gated Geiger mode. Concurrently, the NPG generates short negative 2 MHz pulses that drive a gain-switched distributed feedback laser diode (LD; NEC NX 8563LB). We first measured the performance of the APDs in terms of quantum efficiency (QE) and dark count probability (DCP). Fig. 5 shows the behavior of QE and DCP versus bias voltage for both APDs. When the bias voltage increases above 55 V, both QE and DCP increase gradually. We selected a 10 % QE to set up the APD; for this level of QE, DCP = $4 \times 10^{-5}$ for APD 1, and DCP = $2 \times 10^{-5}$ for APD 2.

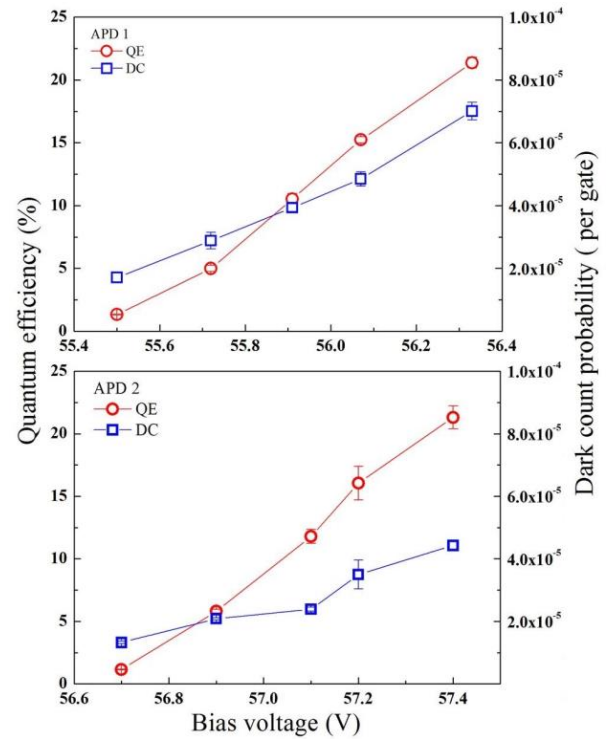

FIG. 5.   Quantum efficiency and dark count probability for APDs 1 and 2.

The repeated laser pulses are attenuated to the range of 0.1–500 photons/pulse through a VOA consisting of passive variable optical attenuators. Each attenuated laser pulse is split into two laser pulses by the BS 50/50, and one of the resulting pulses is delayed through the DL. The two laser pulses (first and second) are sent to Eve–Alice, where the second laser pulse is modulated from 0 to $3\pi/2$ by the PM. The two laser pulses are then reflected by the FM, and the second laser pulse is modulated again by the PM [49]. When the two laser pulses arrive at Bob, the first laser pulse is modulated from 0 to $\pi/2$ by the PM. We use a delay function in the PPG to synchronize the timing of the gate pulses to that of the arriving photons. A thermoelectric cooler control module maintains the temperature at 233 K to prevent temperature variations, which might affect the measurement results. The APD output signals are connected directly to the comparator or the proposed CM scheme, with the balanced APD scheme described in Fig. 3 (not shown in Fig. 4).

TABLE I. APD click information after sifting process.

| Alice PM | | Eve-Bob PM | | Eve-Bob APD click | | Eve-Alice PM | | Bob PM | | Bob APD click | | Type of Case | Graph |
|---|---|---|---|---|---|---|---|---|---|---|---|---|---|
| Phase | Basis | Phase | Basis | Eve-APD 1 (BS) | Eve-APD 2 (Cir) | Phase | Basis | Phase | Basis | APD 1 (BS) | APD 2 (Cir) | | |
| 0 | 0 | 0 | 0 | √ | | 0 | 0 | 0 | 0 | √ | | Case A | (A) |
| | | | | | | | | | | | | Case B | |
| $\pi/2$ | 1 | 0 | 0 | √ | | 0 | 0 | $\pi/2$ | 1 | 50% | 50% | Case C | (B) |
| | | | | | √ | $\pi$ | 0 | $\pi/2$ | 1 | 50% | 50% | Case C | (C) |
| $\pi$ | 0 | 0 | 0 | | √ | $\pi$ | 0 | 0 | 0 | | √ | Case A | (D) |
| | | | | | | | | | | | | Case B | |
| $3\pi/2$ | 1 | 0 | 0 | √ | | 0 | 0 | $\pi/2$ | 1 | 50% | 50% | Case C | (B)` |
| | | | | | √ | $\pi$ | 0 | $\pi/2$ | 1 | 50% | 50% | Case C | (C)` |
| 0 | 0 | $\pi/2$ | 1 | √ | | $\pi/2$ | 1 | 0 | 0 | 50% | 50% | Case C | (E) |
| | | | | | √ | $3\pi/2$ | 1 | 0 | 0 | 50% | 50% | Case C | (F) |
| $\pi/2$ | 1 | $\pi/2$ | 1 | √ | | $\pi/2$ | 1 | $\pi/2$ | 1 | √ | | Case A | (G) |
| | | | | | | | | | | | | Case B | |
| $\pi$ | 0 | $\pi/2$ | 1 | √ | | $\pi/2$ | 1 | 0 | 0 | 50% | 50% | Case C | (F)` |
| | | | | | √ | $3\pi/2$ | 1 | 0 | 0 | 50% | 50% | Case C | (G)` |
| $3\pi/2$ | 1 | $\pi/2$ | 1 | | √ | $3\pi/2$ | 1 | $\pi/2$ | 1 | | √ | Case A | (H) |
| | | | | | | | | | | | | Case B | |

Table I summarizes the APD click information when performing intercept-and-resend methods. After the sifting process, click information using different bases for Alice and Bob has been removed. Only the equal bases cases remain and contribute to generate secure keys or increase QBER (as discussed in Section III). After the sifting process, three cases are possible, concerning the APD click information.

Case A. Let us assume that Alice and Eve–Bob were using the same basis; in this case, only one of the two Eve–Bob APDs generates a click. As a result, Eve can correctly guess Alice's phase information, and modulates that phase in the second laser pulse. If the phase is the same between Eve–Alice and Bob, a click occurs in APD 1. In this case, it is considered that Eve successfully controls Bob's APD click, because Eve and Bob share the same phase information. It also means that, in this case, Eve can obtain all the sifting keys, as a result of her control over Bob's APD clicks.

Case B. Even though an APD click should occur in APD 1 or 2 (depending on Eve-Alice's and Bob's PM), there is no APD click, because of the limited detection efficiency of the APDs. The detection efficiency of an APD is normally in the 1–20 % range.

Case C. If Eve-Bob and Bob have different bases, both Eve–Bob APDs will generate a click with a probability of 50 % (shadowed cells in Table I). This means that Eve may wrongly guess Bob's phase information. Nevertheless, 50 % of the APD clicks in Bob will still generate sifting keys. The other 50 % of the APD clicks will, however, cause an increase in QBER, because Alice and Bob will have different phase information, even though they have the same basis. After the sifting process, the QBER will still remain high, and the increased QBER indicates the existence of Eve.

It may therefore be concluded that case C, when Alice and Bob have the same basis, must be examined carefully, because it is the only case that may lead to an increase in QBER. Before describing in detail how to blind detectors that use a BNC scheme, we present (in Fig. 6) experimental measurement results of APD clicks versus incident photon flux for a conventional QKD system without a BNC scheme. In quantum hacking events such as the one described in Section II, the incident photon flux is one of the most important parameters. All cases in Table I are tested, except for (b)`, (c)`, (f)`, and (g)` because they are exactly the same as (b), (c), (f), and (g). The ideal avalanche count rates are calculated (the calculation details are presented in the Appendix). As expected, if there are no differences between Eve–Bob and Bob PMs (Case A), only one of the two APDs generates an avalanche with a probability of 100 %, as shown in Fig. 6(a), (d), (g), and (h). In these graphs, it is considered that Eve can control the APDs of Bob. Although the count rate of the other APD is also slightly increased owing to the erroneous count rates of optical and electrical devices, this effect is negligible.

In contrast, if Eve–Bob and Bob PMs differ (Case C), the click probability of APD 1 or APD 2 is 50 %, as shown in Fig. 6(b), (c), (e), and (f). In these graphs, both APDs have the same avalanche count rate, which may increase QBER. Therefore, Eve can be detected and its attempt at quantum hacking fails.

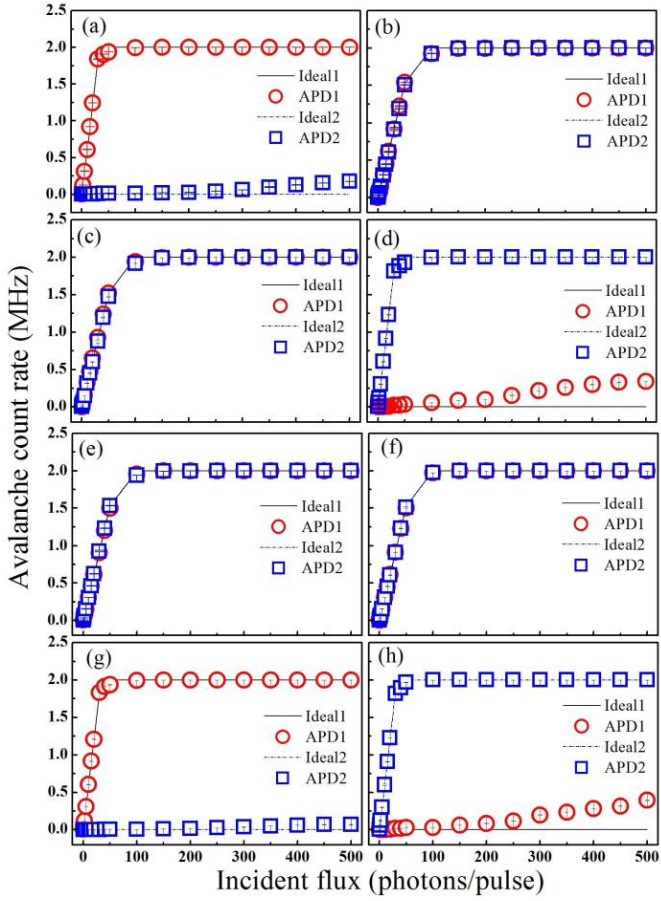

FIG. 6.  Avalanche count rates in the QKD system. Ideal (Ideal 1, 2) and measured count rates (APD1, 2) versus incident flux. Case A is represented by (a), (d), (g), and (h), and Case C by (b), (c), (e), and (f) (see Table I).

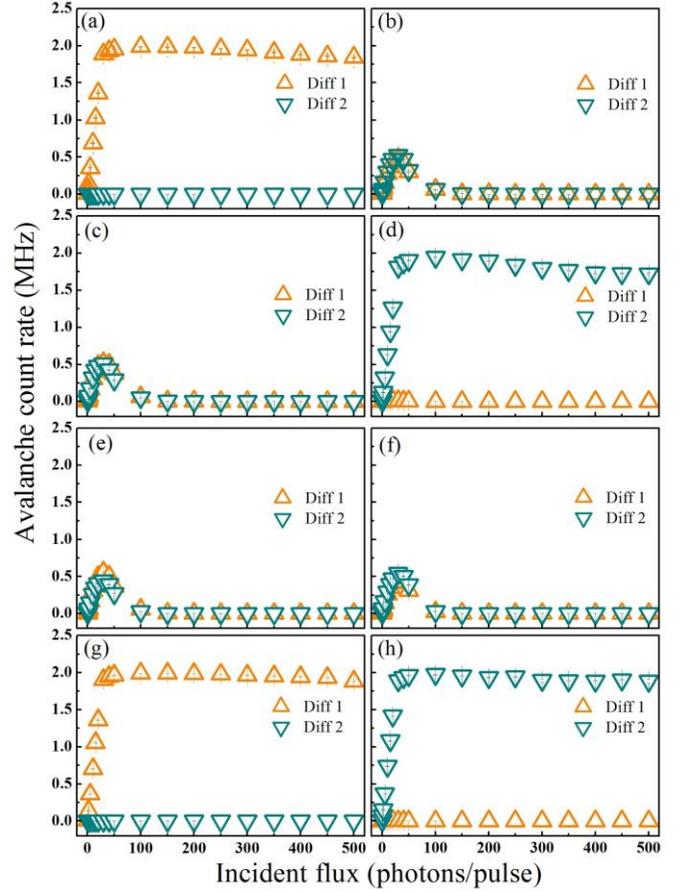

FIG. 7.  Avalanche count rates of the differential amplifier outputs (Diff1, Diff2) and detection of a blinding attack on a plug and play QKD system with a background cancellation scheme. Avalanche count rates versus incident flux. Case A is represented by (a), (d), (g), and (h), and Case C by (b), (c), (e), (f) (see Table I).

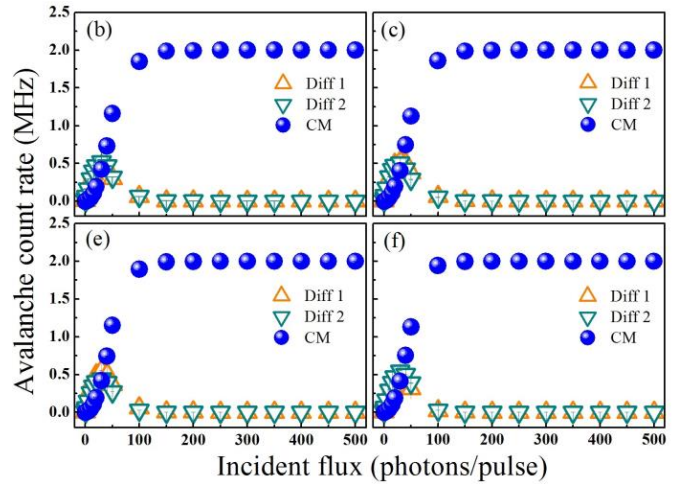

FIG. 8.  Differential amplifier output (Diff1, Diff2) and blinding attack detection using the proposed countermeasure scheme, when Eve attempts a blinding attack on a plug and play QKD system with a background noise cancellation scheme. Avalanche counts rate versus incident flux. All panels correspond to Case C (see Table I).

However, if the QKD adopts a low-noise detector with balanced APDs as described in Section II, the hacking may still succeed, as shown in Fig. 7. Injecting multiple photons blinds the APDs, even if the phase information of Eve–Bob and Bob is different, and the outputs of both APD 1 and 2 come close to zero as the incident photons increase, because of avalanche signal cancellation; this is shown in Fig. 7(b), (c), (e) and (f). This is why any photon coincident photon counts are discarded because we are using two APDs. These graphs show that the avalanche count rate decreases sharply above 100 photons/pulse, because avalanche signals occur at both APDs with almost 100 % of probability. In other words, the detectors become blind completely.

We now evaluate the proposed countermeasure shown in Fig. 3. Fig. 8 shows the avalanche count rates of Diff 1 [Diff amp output (-)] and Diff 2 [Diff amp output (+)] when Eve attempts a blinding attack on the P&P QKD system using balanced APDs shown in Fig. 4. The avalanche count rates of Diff 1 and Diff 2 increase sharply from 0 to 30 photons/pulse, and then decrease to zero above 100 photons/pulse. At the same time, the output signal of the CM increases sharply and its count rate saturates. Clearly, one can easily detect the blinding attack attempt by monitoring the CM count rate. Even though in Case

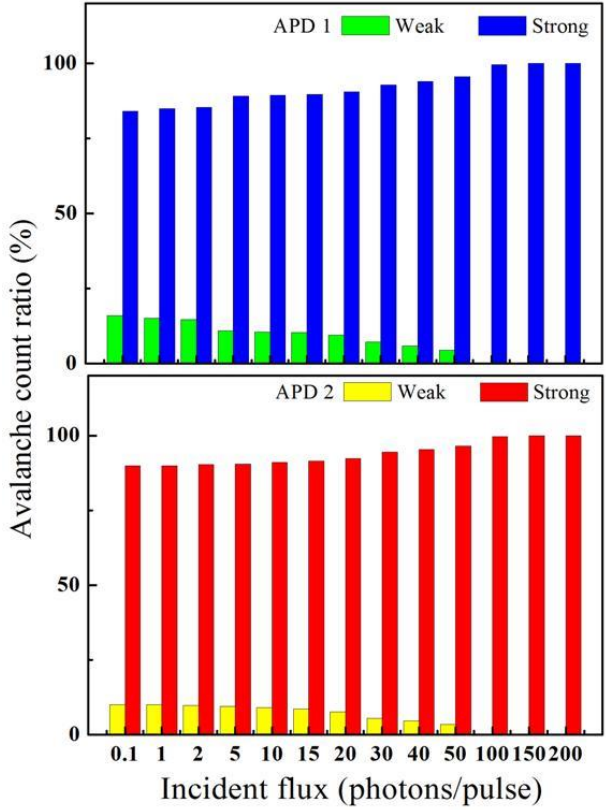

FIG. 9. Weak and strong avalanche count ratios in APD 1 and 2 versus incident flux.

A Eve can control the APDs [as is the case in Fig. 7(a), (d), (g), and (h)], blinding attempts such as those in Fig. 7 (b), (c), (e), and (f) (Case C) are automatically detected. Eve cannot know the exact time when the control operation or blinding operation occurs, because Bob's basis is randomly changed. Therefore, detecting the blinding attempt (and not the controlling operation) is sufficient as a countermeasure.

The proposed countermeasure can only detect strong avalanche signal cancellations. This is sufficient, because at least 100 photons/pulse are required to blind the APDs, as shown in Fig. 7, and furthermore most photons provoke strong avalanches. For example, Fig. 9 shows the avalanche count ratio of the used APDs versus incident flux. As shown, the weak avalanche count ratio that is defined as weak avalanche counts/Total avalanche counts decreases gradually from 0.1 to 50 photons/pulse. In contrast, the strong avalanche count ratio that is defined as strong avalanche counts/Total avalanche counts increases slowly from 0 to 50 photons/pulse, and then saturates at approximately 100 %. This means that above 100 photons/pulse all avalanche signals correspond to strong avalanches.

Fig. 10 indicates the QBER and the CM probability of success versus incident flux. The details on how the QBER caused by the different phase information between Eve-Bob and Bob was calculated are given in the Appendix. The CM probability of success is calculated from the measured data. As shown, the QBER decreases gradually from 0 to 100 photons/pulse, and then stabilizes at 0 % above

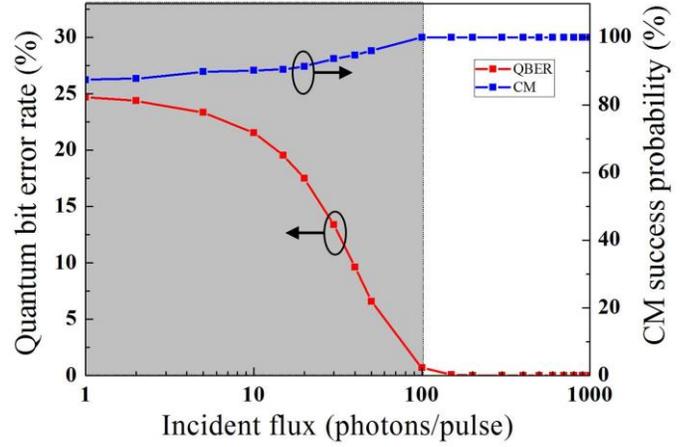

FIG. 10. Quantum bit error rate (QBER) and countermeasure (CM) success probability versus incident flux.

100 photons/pulse. If a single photon arrives at the APD, the QBER will be 25 %. This results from the fact that the total number of sifting key generation cases is 16 (see Table I), the total number of cases typifying Case C is 8, and the APD generates an error count with 50 % probability. Therefore, if a single photon arrives at an APD, an error count will result in 25 % of the cases, on average. The observed decrease in QBER below 100 photons/pulse results from the fact that the flux increase leads to an increase of the probability that both APDs will click. The APDs are becoming therefore increasingly blind and their net output approaches zero. If Eve attempts quantum hacking using a blinding attack, she should therefore use an incident flux higher than 100 photons/pulse, to control the APDs without any increase in QBER. If Eve tries to control using lower intensities than 100 photons/pulse, she will cause an increase in QBER and both Alice and Bob will easily recognize Eve's existence. The CM success probability measures the probability of detecting Eve's blinding attack attempt. It increases slowly from 0 to 100 photons/pulse, and then saturates at 100 % above 100 photons/pulse. As can be seen in Fig. 9, weak avalanche counts account for approximately 10 % of the total avalanche counts at low incident fluxes. Therefore, the CM success probability is approximately 85 % at 1 photon/pulse. However, the weak avalanche counts decrease steadily from 0 to 50 photons/pulse, and then becomes approximately zero at 100 photons/pulse. The CM success probability therefore also becomes 100 % in this range. This means that the proposed CM detects all the blinding attack attempts, and therefore constitutes a perfect protection for QKD systems for this type of attack. It should be noted that, even though the CM success probability is not 100 % for flux values below 100 photons/pulse, this does not imply an unprotected system, given that the QBER will also assist in detecting Eve's presence.

## VI. DISCUSSION

Fig. 11 shows another proposed countermeasure (CM) against quantum hacking of self-differencing APDs. The CM for self-differencing APDs consists of three amplifiers (Amp), three comparators (Comp A, Comp B, and Comp C), and an FPGA. The APD output signal is split into two output signals by splitter 1 (S1). One of the APD output signals is connected to Comp A, which can detect an avalanche signal only when a strong avalanche occurs. The other APD output signal is transferred to splitter 2 (S2), where it is split once more. Next, the differential amplifier subtracts the delayed part from the signal. Finally, the negative and positive outputs of Diff Amp are sent to Comp B and C, respectively. The table inset in Fig. 11 shows the truth table for the logic gate implemented in the FPGA. If Comp A generates a click signal, then a click signal of Comp C should exist in the same time slot and a click signal of Comp B should be generated in the next time slot (because of the delay line in the negative input of the Diff Amp). If there is no signal in Comp B and Comp C with a click signal in Comp A, this means that the APD is blinded ($A\overline{B}\overline{C}$). Therefore, if we measure the logic gate of $A\overline{B}\overline{C}$, we can detect blinding attack attempts from the avalanche signal cancellation. In addition, the proposed CM can detect and distinguish other cases of click information in the APD. If a strong avalanche signal occurs in the APD, the Diff Amp generates an output signal in its positive output port [Fig. 11, pulse (a)]. This case can be detected with a gate implementing the logic expression $A\overline{B}C$; the avalanche signal from pulse (a) will also cause a strong avalanche signal in the delayed APD output, and the Diff Amp will therefore generate an output signal in its negative output port [Fig. 11, pulse (b)]. Even though Comp A and Comp B do not generate any signal in this case, Comp C generates a click signal ($\overline{A}\overline{B}C$), which expresses the fact that a weak avalanche signal occurred in the APD [Fig. 11, pulse (c)]. Even though a weak avalanche signal occurs in the delayed APD output, the Diff Amp generates a click signal in the positive output port, because the weak avalanche signal is removed owing to a blinding signal in the APD output generated by Eve [Fig. 11, pulse (d)]. In the next period, a strong avalanche signal occurs in the APD output. However, the Diff Amp does not generate any signal, because a blinding signal generated by Eve cancels the strong avalanche signal [Fig. 11, pulse (e)]. This case can be detected by a logic gate implementing $A\overline{B}\overline{C}$; a CM output signal indicating a quantum hacking attempt can therefore be generated by the FPGA. We believe that by applying such a CM scheme to QKD systems with self-differencing APDs, these systems can be fully protected from intercept-and-resend methods using APD blinding.

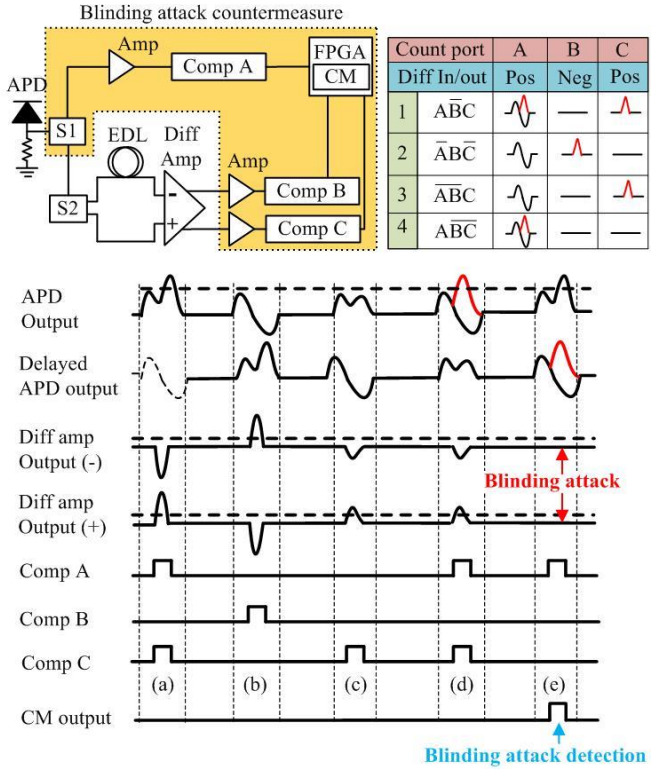

FIG. 11.  Block diagram and timing chart of the countermeasure scheme against blinding attacks in QKD systems using a self-differencing scheme.

## VII. CONCLUSION

We developed a new countermeasure against blinding attacks on low-noise, avalanche photon detectors using a background noise cancellation method in plug and play QKD systems. The effectiveness of the new countermeasure in detecting APD blinding attempts was successfully demonstrated. Our countermeasure scheme can be invaluable in overcoming the security limitations of avalanche photon detectors using a background noise cancellation method.

## APPENDIX

The ideal avalanche count rate when the same phase occurs ($AVC_{One\ click}$) can be expressed by

$$AVC_{One\ click} = \mu_{APD} \times QE \times f_{Gate} , \qquad (1)$$

where $\mu_{APD}$, $QE$, and $f_{Gate}$ are, respectively, the average number of photons per laser pulse in the APD, the APD quantum efficiency, and the frequency of the gate pulses. In contrast, when different phases occur, the ideal avalanche count rate ($AVC_{Both\ clicks}$) is calculated using

$$AVC_{Both\ clicks} = \frac{\mu_{APD} \times QE \times f_{Gate}}{2}. \qquad (2)$$

The QE is selected as 10 % (from Fig. 6) and $f_{Gate}$ = 2 MHz, as generated by the PPG. The average number of photons per laser pulse in the APDs ($\mu_{APD}$) is calculated by

$$\mu_{APD} = \frac{n}{10^{\left(\frac{Att_{Total}}{10}\right)}}, \qquad (3)$$

where $n$ and $Att_{Total}$ are the number of photons in the laser pulses and the total attenuation ratio in the QKD system, respectively. The number of photons in the laser pulses ($n$) can be expressed by

$$n = \frac{E_{Pulse}}{E_{Photon}} \qquad (4)$$

where $E_{Pulse}$ is the total energy of the laser pulse and $E_{Photon}$ is the energy of a single photon of 0.7999 eV at 1.55 μm. The value of $E_{Pulse}$ can be obtained by

$$E_{Pulses} = \frac{P_{Ave}}{R},  \quad (5)$$

where $P_{Ave}$ and $R$ are the average optical power of the laser pulses and their repetition rate, respectively. The value of $P_{Ave}$ can be measured by an optical power meter; $R$ is 2 MHz, as generated by an NPG. We can calculate $n$ using Eq. (4) and Eq. (5).

The value of $Att_{Total}$ is calculated by

$$Att_{Total} = Att_{Bob\ to\ Eve\ Alice} + Att_{Bob},  \quad (6)$$

where $Att_{Bob\ to\ Eve\ Alice}$ and $Att_{Bob}$ are the attenuation ratio from Bob to Eve–Alice, and the attenuation ratio in Bob, respectively. The value of $Att_{Bob}$ can be measured by an optical power meter; the value of $Att_{Bob\ to\ Eve\ Alice}$ can be obtained by

$$Att_{Bob\ to\ Eve\ Alice} = 10 \times log\left(\frac{n}{\mu_{Eve\ Alice}}\right),  \quad (7)$$

where $\mu_{Eve\ Alice}$ is the average number of photons per laser pulse in Eve–Alice. The average number of photons is determined by the attenuation ratio in the VOA. We can calculate the average number of photons per laser pulse in the APDs ($\mu_{APD}$) using Eq. (4) and Eq. (6). Having done that, and using Eq. (1) and Eq. (2), we can calculate the ideal avalanche count rate (Ideal 1 and 2 in Fig. 6).

When a laser pulse with an average number of photons per pulse μ arrives at the APD, we may consider two click probabilities: the single click probability $p_1(\mu)$, and the double click probability $p_2(\mu)$:

$$p_1(\mu) \rightarrow QBER_1 = \frac{1}{2},  \quad (8)$$

$$p_2(\mu) \rightarrow QBER_2 = 0.  \quad (9)$$

Applying these probabilities to a QKD system with a total number of $N$ laser pulses, we obtain:
- When Eve-Bob and Bob have the same basis,

$$\frac{N}{2} \times p_2(\mu) \rightarrow QBER = 0.  \quad (10)$$

- When Eve-Bob and Bob have different bases,

$$Single\ click = \frac{N}{2} \times p_1(\mu) \rightarrow QBER  \quad (11)$$

$$Double\ click = \frac{N}{2} \times p_2(\mu) \rightarrow QBER = 0.  \quad (12)$$

The QBER caused by the different phase ($QBER_{Diff\ phase}$) can be expressed by

$$QBER_{Diff\ phase} = \frac{\frac{N}{2} \times p_1(\mu) \times \frac{1}{2}}{\frac{N}{2} \times p_S(\mu) + \frac{N}{2} \times p_1(\mu)}.  \quad (13)$$

The CM success probability ($p_{CM}$) can therefore be expressed by:

$$p_{CM} = \frac{Strong\ avalanche\ counts}{Weak\ avalanche\ counts + Stron\ avalanche\ counts} \times 100.  \quad (14)$$

## REFERENCES

[1] C. H. Bennett and G. Brassard, *in Proceedings of the International Conference Computers, Systems and Signal Processing*, 1984, p. 175–179.

[2] A. Ekert, Phys. Rev. Lett., **67**, 661 (1991).

[3] C. H. Bennett, Phys. Rev. Lett., **68**, 3121 (1992).

[4] A. Muller, T. Herzog, B. Huttner, W. Tittel, H. Zbinden, and N. Gisin, Appl. Phys. Lett., **70**, 793 (1997).

[5] D. Stucki, N. Gisin, O. Guinnard, G. Ribordy, H. Zbinden, New J. Phys., **4**, 41.1 (2002).

[6] H. K. Lo, X. F. Ma, and K. Chen, Phys. Rev. Lett., **94**, 230504-1 (2005).

[7] Z.L. Yuan, B.E. Kardynal, A.W. Sharpe, and A.J. Shields, Appl. Phys. Lett., **91**, 041114-1104114-3 (2007).

[8] B. E. Kardynal, Z. L. Yuan, and A. J. Shields, Nat. Photonics, **2**, 425 (2008).

[9] J. F. Dynes, Z. L. Yuan, A. W. Sharpe, and A. J. Shields, Appl. Phys. Lett., **93**, 031109-1 (2008).

[10] A. R. Dixon, Z. L. Yuan, J. F. Dynes, A. W. Sharpe, and A. J. Shields, Opt. Express, **16**, 23, 18790 (2008).

[11] A. Laing, V. Scarani, J. G. Rarity, and J. L. O'Brien, Phys. Rev. A, **82**, 012304-1 (2010).

[12] A. R. Dixon, Z. L. Yuan, J. F. Dynes, A. W. Sharpe, and A. J. Shields, Appl. Phys. Lett., **96**, 16, 161102-1 (2010).

[13] H.-K. Lo, M. Curty, and B. Qi, Phys. Rev. Lett., **108**, 130503-1 (2012).

[14] P. D. Townsend, Nature, **385**, 47 (1997).

[15] W. Chen *et al*, IEEE Photon. Technol. Lett., **21**, 575 (2009).

[16] T.-Y. Chen *et al.*, Opt. Express, **18**, 26, 27217 (2010).

[17] S. Wang *et al.*, Opt. Lett., **35**, 14, 2454 (2010).

[18] M. Sasaki *et al.*, Opt. Express, **19**, 11, 10387 (2011).

[19] D. Stucki *et al.*, New J. Phys. **13**, 12, 123001-1 (2011).

[20] P. Jouguet *et al.*, Opt. Express, **20**, 13, 14030 (2012).

[21] K. Yoshino, T. Ochi, M. Fujiwara, M. Sasaki, and A. Tajima, Opt. Express, **21**, 25, 31395 (2013).

[22] K. Shimizu, T. Honjo, M. Fujiwara, T. Ito, K. Tamaki, S. Miki, T. Yamashita, H. Terai, Z. Wang, and M. Sasaki, J. Lightwave Technol., **32**, 1, 141 (2014).

[23] S. Wang *et al.*, Opt. Express, **22**, 18, 21739 (2014).

[24] A. R. Dixon *et al.*, Opt. Express, **23**, 6, 7583 (2015).

[25] H.-F. Zhang, J. Wang, K. Cui, C.-L. Luo, S.-Z. Lin, L. Zhou, H. Liang, T.-Y. Chen, K. Chen, and J.-W. Pan, J. Lightwave Technol., **30**, 20, 3226 (2012).

[26] H. K. Lo, M. Curty, and K.Tamaki, Nature Photon., **8**, 595 (2014).

[27] L. Lydersen, C. Wiechers, C. Wittmann, D. Elser, J. Skaar, and V. Makarov, Nature Photon., **4**, 686 (2010).

[28] Z. L. Yuan, J. F. Dynes, and A. J. Shields, Nature Photon., **4**, 800 (2010).

[29] L. Lydersen, C. Wiechers, C. Wittmann, D. Elser, J. Skaar, and V. Makarov, Nature Photon., **4**, 801 (2010).

[30] I. Gerhardt, Q. Liu, A. Lamas-Linares, J. Skaar, C. Kurtsiefer, and V. Makarov, Nature Commun., **2**, 1 (2011).

[31] B. Qi, C.-H. F. Fung, H.-K. Lo, and X. Ma, Quant. Inf. Comp., **7**, 73 (2007).

[32] Y. Zhao, C.-H. F. Fung, B. Qi, C. Chen, and H.-K. Lo, Phys. Rev. A, **78**, 042333-1 (2008).

[33] V. Makarov, A. Anisimov, and J. Skaar, Phys. Rev. A, **74**, 022313-1 (2006).

[34] V. Makarov, A. Anisimov, and J. Skaar, Phys. Rev. A, **78**, 019905-1 (2008).

[35] H. Weier, H. Krauss, M. Rau, M. Fürst, S. Nauerth, and H. Weinfurter, New J. Phys., **13**, 1 (2011).

[36] F. Xu, B. Qi, and H.-K. Lo, New J. Phys., **12**, 1 (2010).

[37] N. Gisin, S. Fasel, B. Kraus, H. Zbinden, and G. Ribordy, Phys. Rev. A, **73**, 022320-1 (2006).

[38] S.-H. Sun, M.-S. Jiang, and L.-M. Liang, Phys. Rev. A, **83**, 062331-1 (2011).

[39] J. Z. Huang, C. Weedbrook, Z. Q. Yin, S. Wang, H. W. Li, W. Chen, G. -C. Guo, and Z. F. Han, Phys. Rev. A, **87**, 062329-1 (2013).

[40] Y. L. Tang, H. L. Yin, X. Ma, C. H. F. Fung, Y. Liu, H. L. Yong, T.–Y. Chen, C.–Z. Peng, Z.–B. Chen, and J. W. Pan, Phys. Rev. A, **88**, 022308-1 (2013).

[41] P. Jouguet, S. Kunz-Jacques, and E. Diamanti, Phys. Rev. A, **87**, 062313-1 (2013).

[42] Z. L. Yuan, A. W. Sharpe, J. F. Dynes, A. R. Dixon, and A. J. Shields, Appl. Phys. Lett., **96**, 071101-1 (2010).

[43] K. A. Patel, J. F. Dynes, A. W. Sharpe, Z. L. Yuan, R. V. Penty, and A. J. Shields, Electron. Lett., **48**, 111 (2012).

[44] J. Zhang, M. A. Itzler, H. Zbinden, and J. W. Pan, Light Sci. Appl., **4**, 1 (2015).

[45] L. C. Comandar, B. Fröhlich, M. Lucamarini, K. A. Patel, A. W. Sharpe, J. F. Dynes, Z. L. Yuan, R. V. Penty, and A. J. Shields, Appl. Phys. Lett., **104**, 021101-1 (2014).

[46] L. C. Comandar, B. Fröhlich, J. F. Dynes, A. W. Sharpe, M. Lucamarini, Z. L. Yuan, R. V. Penty, A. J. Shields, J. Appl. Phys., **117**, 083109-1 (2015).

[47] A. Tomita and K. Nakamura, Optics Lett., **27**, 1827 (2002).

[48] M. S. Jiang, S. H. Sun, G. Z. Tang, X. C. Ma, C. Y. Li, and L. M. Liang, Phys. Rev. A, **88**, 062335-1 (2013).

[49] O. Kwon, M. S. Lee, M. K. Woo, B. K. Park, I Y. Kim, Y. S. Kim, S. W. Han, and S. Moon, Laser Phys., **25**, 1–5 (2015).